\documentstyle[11pt,aaspp4,flushrt,tighten,psfig]{article}  

\begin{document}

\title{First Results from the X--ray and Optical\footnote{Based on
observations performed at the European Southern Observatory, Paranal,
Chile} Survey of the Chandra  Deep Field South} 
\author{R. Giacconi\altaffilmark{1,2}, P. Rosati \altaffilmark{3}, P. 
Tozzi\altaffilmark{1,4}, M. Nonino\altaffilmark{4}, G. 
Hasinger\altaffilmark{5}, C. Norman\altaffilmark{1,6}, J. 
Bergeron\altaffilmark{3},
S. Borgani\altaffilmark{7}, R. Gilli\altaffilmark{1,8}, R. 
Gilmozzi\altaffilmark{3}, and W. Zheng\altaffilmark{1}}

\affil{$^1$Dept. of Physics and Astronomy, The
Johns Hopkins University, Baltimore, MD 21218, USA}
\affil{$^2$Associated Universities Inc., 1400, 16th st. NW, Washington
DC 20036, USA} 
\affil{$^3$European Southern Observatory,
Karl-Schwarzschild-Strasse 2, D-85748 Garching, Germany}
\affil{$^4$Osservatorio
Astronomico di Trieste, via G.B. Tiepolo 11, I--34131, Trieste, Italy}
\affil{$^5$Astrophysikalisches Institut,
An der Sternwarte 16, Potsdam 14482 Germany} 
\affil{$^6$Space Telescope Science Institute, 3700 S. Martin Drive,
Baltimore, MD 21210, USA}
\affil{$^7$INFN, c/o Dip. di Astronomia dell'Universit\`a, via Tiepolo
11, I--34131, Trieste, Italy} 
\affil{$^8$Universit\`a degli Studi di Firenze, Dip. Astronomia, Largo
E. Fermi 5, I-50125, Firenze, Italy}

\begin{abstract}
We present our first results from 130 ks of X--ray observations
obtained with the Advanced CCD Imaging Spectrometer on the Chandra
X--ray Observatory.  The field of the two combined exposures is 0.096
square degrees and the detection limit is to a $S/N$ ratio of 2
(corresponding to $\sim 7$ net counts).  We reach a flux of $2 \times
10^{-16}$ erg s$^{-1}$ cm$^{-2}$ in the 0.5--2 keV soft band and $2
\times 10^{-15}$ erg s$^{-1}$ cm$^{-2}$ in the 2--10 keV hard band.
Our combined sample has 144 soft sources and 91 hard sources
respectively for a total of 159 sources.  Fifteen sources are detected
only in the hard band, and 68 only in the soft band.

For the optical identification we carried out a survey in VRI with the
FORS--1 imaging-spectrometer on the ANTU telescope (UT--1 at VLT)
complete to R$\le 26$. This dataset was complemented with data from
the ESO Imaging Survey (EIS) in the UBJK bands and the ESO Wide Field
Imager Survey (WFI) in the B band.  The positional accuracy of the
X--ray detections is of order of 1'' in the central 6'. Optical
identifications are found for $\simeq 90$\% of the sources. Optical
spectra have been obtained for 12 objects.

We obtain the cumulative spectra of the faint and bright X--ray
sources in the sample and also the hardness ratios of individual
sources.  A power law fit in the range 2--10 keV using the galactic
value of $N_H \simeq 8 \times 10^{19}$ cm$^{-2}$, yields a photon
index of $ \Gamma = 1.70 \pm 0.12$ and $1.35 \pm 0.20$ (errors at 90\%
c.l.) for the bright and faint sample respectively, showing a
flattening of the spectrum at lower fluxes. Hardness ratio is given as
a function of X--ray flux and confirms this result. The spectrum of
our sources is approaching the spectrum of the XRB in the hard band,
which has an effective $\Gamma = 1.4$.

Correlation function analysis for the angular distribution of the
sources indicates that they are significantly clustered on scales as
large as 100 arcsec. The scale-dependence of the correlation function
is a power law with index $\gamma \sim 2$, consistent with that of the
galaxy distribution in the local Universe. Consequently, the discrete
sources detected by deep Chandra pointed observations can be used as
powerful tracers of the large-scale structure at high redshift.

We discuss the LogN--LogS relationship and the discrete source
contribution to the integrated X--ray sky flux.  In the soft band, the
sources detected in the field at fluxes below $10^{-15}$ erg s$^{-1}$
cm$^{-2}$ contribute $(4.0 \pm 0.3) \times 10^{-12}$ erg cm$^{-2}$
s$^{-1}$ deg$^{-2}$ to the total XRB.  The flux resolved in the hard
band down to the flux limit of $2 \times 10^{-15}$ erg s$^{-1}$
cm$^{-2}$, contributes $ (1.05 \pm 0.2) \times 10^{-11}$ erg cm$^{-2}$
s$^{-1}$ deg$^{-2}$.  Once the contribution from the bright counts
resolved by ASCA is included, the total resolved XRB amounts to $1.3
\times 10^{-11}$ erg cm$^{-2}$ s$^{-1}$ deg$^{-2}$ which is a
fraction of 60--80\% of the total measured background.  This result
confirms that the XRB is due to the integrated contribution of
discrete sources, but shows that there is still a relevant fraction
(at least 20\%) of the hard XRB to be resolved at fluxes below
$10^{-15}$ erg s$^{-1}$ cm$^{-2}$.

We discuss the X--ray flux versus R magnitude relation for the
identified sources.  We find that $\simeq 10\%$ of the sources in our
sample are not immediately identifiable at $R>26$.  For these sources
$S_X/S_{opt} > 15 $, whereas most of the ROSAT and Chandra sources
have $S_X/S_{opt} < 10$.  We have found also a population of objects
with unusually low $S_X/S_{opt}$ that are identified as galaxies.  The
R--K vs R color diagram shows that the Chandra sources continue the
trend seen by ROSAT.

For our 12 spectroscopically studied objects with redshifts, we
observe 4 QSOs, 5 Sy2 galaxies, 1 elliptical and 2 interacting
galaxies.  We compare the $L_X$ vs $z$ obtained with these
measurements and show that Chandra is achieving the predicted
sensitivity.
\end{abstract}


\newpage

\section{INTRODUCTION}

In the rocket flight which discovered X-ray stars (Giacconi et al
1962), the presence of a diffused X-ray background (XRB) was also
observed.  Reviews of the findings up to the middle 70's are given in
Gursky and Schwartz (1977), Fabian and Barcons (1992) and references
therein.  The measurement by Uhuru of the high degree of isotropy led
to the conclusion that the XRB had to be extragalactic.  The
logN--logS relation for sources at high galactic latitude ($b>20$)
showed a slope of 1.5 which would account for the entire XRB already
at fluxes larger than 10$^{-14}$ erg cm$^{-2}$ s$^{-1}$ (Matilsky et
al 1973).  The required number of discrete sources had to be large ($N
> 10^6$ sr$^{-1}$) to be consistent with the value of the fluctuations
of $< 3$\% over 2 degrees (Schwartz et al 1976).

The spectrum of the XRB had been measured over the range 3 to 100 keV
and could be fitted with two power laws of index $\Gamma$ = 1.4 for $E
< 25$ keV and $\Gamma = 2.4$ for $E > 25$ keV (Gursky and Schwartz
1977).  However, the results of HEAO--1 gave an excellent fit over the
$3-30$ keV range of the XRB spectrum to a thermal bremstrahlung from
a hot plasma at 40 keV, leading the authors (Marshall et al 1980) to
suggest that the flux could be due to a hot intergalactic medium as
had been considered by Field and Perrenod (1977).  The finding that
the spectrum of AGNs had been measured to be steeper than that of the
XRB (Mushotzky 1984) further strengthened the view that the XRB could
not be due to the sum of known sources.

On the other hand, the Einstein survey showed that about 25\% of the
soft XRB (1--3 keV) was resolved into discrete sources at a flux of
order $3 \times 10^{-14}$ erg s$^{-1}$ cm$^{-2}$ (Giacconi 1980,
Griffiths et al. 1983).  A large fraction of these sources were AGNs.
A reasonable extrapolation of the X-ray properties and optical counts
of extragalactic sources led to the conclusion that it was unlikely
that discrete sources contributed less than 50\% of the XRB (e.g.,
Schmidt \& Green 1986; Setti 1985).  If so, and their spectrum
continued to be steep (energy index $ \alpha \sim 0.7-1.2$), then one
could show that a diffuse bremsstrahlung hypothesis for the residual
was untenable (Giacconi \& Zamorani 1987).  The COBE result entirely
ruled out the possibility that the XRB could be due to thermal
bremsstrahlung from a hot plasma (Mather et al. 1990).

The ROSAT survey established that 70--80\% of the XRB is resolved into
discrete sources in the 1-2 keV range at a flux level of $1 \times
10^{-15}$ erg s$^{-1}$ cm$^{-2}$ (Hasinger et al. 1998).  At higher
energies (2--10 keV) the ASCA and BeppoSAX satellites have achieved
detection limits of $3-5 \times 10^{-14}$ erg s$^{-1}$ cm$^{-2}$ and
have resolved $\sim 25\%$ of the 2-10 keV XRB into discrete sources
(Ogasaka 1998; Cagnoni et al. 1998; Della Ceca et al. 1999a; Giommi,
Fiore \& Perri 1998).  The population of sources appears to be
composed of obscured and unobscured AGNs.  The majority of hard
sources appear to have counterparts in the soft X--ray band and are
believed to be obscured AGNs.  In this case, the soft emission may be
due to electron scattered radiation.  In addition to the scattered
radiation, there are in principle several possibilities why obscured
AGN could show up in the soft band: a very high redshift, so that the
absorption cutoff is shifted to the soft energy band; a starburst
component; photoionized gas emission (as observed by XMM in NGC 1068,
Paerels 2000).

The very high sensitivity, broad energy range and high angular
resolution of the Chandra observatory (Weisskopf, O'dell \& van
Speybroeck 1996) will permit the final resolution of the origin of the
XRB at least in the 0.5 to 10 keV range.  However, the most important
aspect of the deep surveys (exposure time $\simeq 1$ Msec) to be
conducted with Chandra will be the study of the individual classes of
objects rather than their integrated properties.  Based on the
limiting sensitivity, exposure map and source surface density found in
the first Chandra data, we expect a limiting flux of $f_{[0.5-2 keV]}
= 2\times 10^{-17}$erg s$^{-1}$ cm$^{-2}$ in the center of the field
for 1.5 Msec of exposure.  Such a depth will permit the study of: (1)
the formation and evolution of AGNs at large redshifts ($z \simeq 5$);
(2) clusters of galaxies at redshifts greater than $2$ and (3) study
of star--forming regions in galaxies at moderate redshifts ($z \simeq
1$).

We selected a field for our observations which has the following
properties: (1) very low Galactic neutral hydrogen column ($N_H \simeq
8 \times 10^{19} \rm cm^{-2}$) -- a Southern sky equivalent to the
Lockman Hole; (2) no stars brighter than $m_v=14$ and (3) well suited
to observations with 8 meter class telescopes such as the VLT and
Gemini.  Our observations are centered at RA = 3:32:28.0 DEC =
-27:48:30 (J2000).  The results in this paper pertain to the initial
130 ksec of observations out of the currently planned 500 ksec
(another exposure of 500 ksec has been approved, for a total of 1
Msec, for the end of the year 2000).  The plan of the paper is as
follows.  In \S 2 we describe the data reduction.  In \S3 we will
present the results on the X--ray data including spectral properties
of the sources, number counts and correlation function.  In \S 4 we
will describe the optical data.  In \S 5 we will discuss our results.
Finally, we will summarize our conclusions in \S 6.

\section{X--RAY DATA: OBSERVATIONS AND DATA REDUCTION}

The Chandra Deep Field South (CDFS) data are obtained from the
combination of two exposures taken when the Advanced CCD Imaging
Spectrometer--Imaging (ACIS-I) detector was at a temperature of -110
K.  The first observation (Obs ID 1431\_0) was taken on 1999-10-15/16,
in the very faint mode, for a total 34 ks exposure. The second
observation (Obs ID 1431\_1) was taken on 1999-11-23/24, in the faint
mode, for a total 100 ks exposure.

The data were reduced and analyzed using the CIAO software (release
V1.3, see http://asc.harvard.edu/cda).  The Level 1 data were
processed using the latest calibration files and the latest aspect
solution. We included the new quantum efficiency uniformity files,
that correct the effective area for loss due to charge transfer
inefficiency at a temperature of -110 K.  This correction compensates
for the loss of events far from the readout, especially at high
energies, and it is particularly relevant when fitting the total
spectrum in the broadest energy range.

The data were filtered to include only the standard event grades
0,2,3,4 and 6.  All hot pixels and columns were removed as were also
the columns close to the border of each node, since the grade
filtering is not efficient in these columns.  We looked for flickering
pixels, defined as pixels with more than 4 events contiguous in
time. The removal of columns and pixels reduces slightly the effective
area of the detector and this effect has been included when computing
the total exposure maps.  Time intervals with background rates larger
than $3$ sigma over the quiescent value ($0.31$ counts s$^{-1}$ per
chip in the 0.3--10 keV band) were removed.  This procedure gave 25
ksec of effective exposure out of the first observation, and 93 ksec out
of the second, for a total of 118 ks.  The two observations had
different roll angles, so that the exposure in the combined image of
the field ranges between 25 and 118 ksec over a total coverage of 0.096
deg$^2$.  Since the two observations have the same nominal aimpoint,
the exposure time is 118 ksec over the majority of the field of view.
The chip S--2 was active during both observations. We did not include
data from this chip in the following analysis, since the point spread
function (PSF) is very broadened ($\simeq 15$ arcsec) and it adds a
very small effective area especially at low fluxes.

We selected a soft band from 0.5--2 keV and a hard band from 2--7 keV in
which to detect sources.  The hard band was cut at 7 keV since above
this energy the effective area of Chandra is decreasing, while the
instrumental background is rising, giving a very inefficient detection
of sky photons.  A wavelet detection algorithm run on the 7--10 keV
image, gives no sources over our detection threshold. Note also that
the counts expected on the ACIS-I detector in the 7--10 keV band for the
hardest sources, are always no more than 3\% of the total hard
photons. Thus, including the 7--10 keV band would always decrease the
signal for the hard sources. However, we will always quote the fluxes in
the canonical 2--10 keV band, as extrapolated from the flux measured
from the 2--7 keV count rate, in order to have a direct comparison with
the previous results.

Figure 1 shows a sky map of the field.  We used a wavelet detection
algorithm (Rosati et al. 1995) to find X--ray sources.  In order to
match the PSF variation as a function of the off-axis angle, the
wavelet analysis was carried out at five scales, $a_i=(\sqrt{2})^i$
pixels, for $i=0,1,2,3,4$ and pixel size 0.984 arcsec.  Simulations
have shown that simple aperture photometry is very accurate across the
ACIS--I detector and was therefore preferred to time--consuming
wavelet photometry (see Rosati et al. 1995).  The area of extraction
of each source is defined as a circle of radius $R_s=2.4\times FWHM$
(with a minimum of 5 pixels $(\simeq 5\arcsec$).  The FWHM (in arcsec)
is modeled as a function of the off-axis angle $\theta$ (in arcmin) as
$FWHM(arcsec)=\sum_{i=0,3} a_i\theta^i$, with
$a_i=\{0.678,-0.0405,0.0535\}$ (see the Observatory Guide,
http://asc.harvard.edu/udocs/docs/docs.html).  The background was
calculated locally for each source in an annulus with outer radius of
$R_s+8''$ and inner radius of $R_s+2''$, after masking out other
sources.  With this choice, the average number of counts in the
background regions is $\simeq$ 11--22 in the soft band and $ \simeq$
17--34 in the hard band.  Thus the local estimate of the background
has a poissonian fluctuation always less than 30\%.  We define the
$S/N$ ratio as $S/\sqrt{S+2B}$ where S are the net counts in the
extraction region of radius $R_s$, and B are the background counts
found in the annulus defined above and rescaled to the extraction
region.  Since the lowest $S/N$ ratio of our final catalog is 2 across
the field, the flux threshold across the field in the soft band,
including the effect of vignetting and of the point spread function,
is $ \simeq 2\times 10^{-16} $ erg s$^{-1}$ cm$^{-2}$ within the
central 6 arcmin, and grows to $\simeq 5\times 10^{-16} $ erg s$^{-1}$
cm$^{-2}$ at 10 arcmin off--axis.  In the hard band, the flux limit is
$ \simeq 2\times 10^{-15} $ erg s$^{-1}$ cm$^{-2}$ within the central
6 arcmin, and $\simeq 4\times 10^{-15} $ erg s$^{-1}$ cm$^{-2}$ 10
arcmin off--axis.  In the following we describe the analysis of
point--like sources only, leaving the analysis of diffuse sources to a
subsequent paper.

We performed extensive simulations to determine the accuracy of our
aperture photometry and our completeness limit.  The aperure
photometry with the aforementioned parameters is accurate within 3\%.
From the catalog of candidate sources selected by the wavelet
detector, we remove all sources with a signal-to-noise ratio $S/N<2$,
corresponding to 7 net counts in the center.  With these criteria, our
detections have a less than 5\% probability of including a fake source
on the total ACIS-I field.  This is a robust limit since the
background is very low and we are always signal limited.  More
importantly, the simulations provide a test for the sky coverage
model, which is defined as the area of the sky where a point--like
source of a given flux can be detected by the wavelet algorithm and
has a $S/N >2$ in the extraction region of radius $R_S$. The effective
solid angle of the Chandra observations is equal to the geometrical
solid angle (0.096 deg$^2$) only at fluxes $>2\times 10^{-15}, 2\times
10^{-14}$ erg s$^{-1}$ cm$^{-2}$ in the soft and hard bands
respectively, whereas it drops at 50\% of these values at fluxes
$<3.2\times 10^{-16},2.3\times 10^{-15}$ erg s$^{-1}$ cm$^{-2}$. By
comparing the input LogN--LogS with the output LogN--LogS in the
simulations we verified that our model for the sky coverage is
accurate within 5\%.  A further check comes from the preliminary
analysis of a total exposure of 300 ksec of the CDFS.  We found the
presence of 5\% of spurious sources, to be added to six sources
associated with flickering pixels that were not removed from the
exposure. 

We used two separate conversion factors to derive the energy flux from
the observed count rate for the soft and hard bands.  The conversion
factors were $(4.8\pm 0.3) \times 10^{-12}$ erg s$^{-1}$ cm$^{-2}$ per
count s$^{-1}$ in the soft band, and $(2.7\pm 0.3) \times 10^{-11}$
erg s$^{-1}$ cm$^{-2}$ per count $\rm s^{-1}$ in the hard band
assuming an absorbing column of $8 \times 10^{19}$ cm$^{-2}$ (Galactic
value) and a photon index, $\Gamma = 1.7$, consistent with the average
spectrum of the bright sample.  The uncertainties in the conversion
factors reflect the range of possible values for the photon index,
$\Gamma = 1.4 - 2.0$.  As suggested by the spectral analysis of the
stacked spectra, these values are representative of our sample (see \S
3.1).  The conversion factors were computed at the aimpoint (which is
the same for both exposures).  Before being applied to the net count
rate of a given sources, the conversion factors are corrected for
vignetting.  The correction is given by the ratio of the value of the
exposure map at the aimpoint to the value of the exposure map at the
source position (averaged over the extraction region).  This is done
separately for the soft and the hard band, using the exposure maps
computed for energies of 1.5 keV (soft) and 4.5 keV (hard).  Sources
with more than 100 total counts in the two bands were analyzed also
with XSPEC (Arnaud 1996), fitting an absorbed power law to the data
binned with a minimum of 20 photons per bin.  The soft and hard fluxes
of the fitted spectra are always within 10\% of the values obtained
using the conversion factors quoted above.

\section{X--RAY DATA: RESULTS}

\subsection{X--ray Spectra and Hardness Ratio}

The majority of the sources are faint. Thus we measure the average
stacked spectrum of both the faint sources and bright sources. To this
purpose, we use only the longest exposure (93 ksec), in order to have
a uniform exposure for the stacked spectra.  From our sample of 159
sources, 5 were excluded because they were present only in the field
of view of the first exposure.  We divided the remaining sample into
two groups of 29 and 125 with a dividing count rate of $1.0 \times
10^{-3}$ cts/sec in the total band 0.5--7 keV.  The background
spectrum was obtained using the
event file of the total field of the same
exposure, after the removal of the detected sources.  The background
is scaled by the ratio of the total exposure maps of the sources and
of the background.  Such a procedure guarantees a correct background
subtraction despite the non-uniformities of the instrumental
background across ACIS-I.  The ancillary response matrix for the
effective area is obtained from the counts--weighted average of the
matrices of the single sources.  The response matrix is assumed to be
the one computed in the aimpoint.  We recall that we use the quantum
efficiency uniformity file appropriate for a temperature of -110 K.
This file corrects the quantum efficiency with respect to the
pre--flight values; especially, it takes into account the correction
due to the charge transfer inefficiency.

We used XSPEC to compute the slope of a power law spectrum with $N_H$
absorption at low energy, in the energy range 0.5--10 keV and in the
2--10 keV band only.  The value of $N_H$ is fixed at the galactic
value when the fit is done using the hard energy range, while, if the
total energy range is used, $N_H$ is left free to vary.  We exclude
bins below $0.5$ keV because the calibration is still uncertain below
this energy.

For the total sample, including all the 154 sources on the largest
exposure, we performed a power-law fit over the energy range 0.5--10
keV, and obtained a photon index $\Gamma$ of $1.53 \pm 0.07$ and a
column density $N_H = (6.0\pm 3.0) \times 10^{20}$ cm$^{-2}$. Errors
refer to the 90\% confidence level. Then we perform the fit using only
the 2--10 keV energy range, with Galactic $N_H = 8 \times 10^{19}$
cm$^{-2}$. We obtained $\Gamma = 1.61\pm 0.11$.  The contribution of
the total sample to the XRB is $ (4.0 \pm 0.3) \times 10^{-12}$ erg
cm$^{-2}$ s$^{-1}$ deg$^{-2}$ in the soft band and $ (1.07 \pm 0.15)
\times 10^{-11}$ erg cm$^{-2}$ s$^{-1}$ deg$^{-2}$ in the hard band.

For the bright sample we obtained a photon index $\Gamma$ of $1.71 \pm
0.07$ and $N_H = (7.0 \pm 2.0) \times 10^{20}$ cm$^{-2}$ from the
fit in the 0.5--10 keV energy range.  Using the 2--10 keV energy range
with galactic $N_H$ we obtained $\Gamma$ of $1.70 \pm 0.12$.  The
contribution of the bright sample to the XRB is $ (6.3 \pm 0.9) \times
10^{-12}$ erg cm$^{-2}$ s$^{-1}$ deg$^{-2}$ in the hard band. 

We note that our bright sample shows a spectrum that is softer than that
measured at about the same fluxes by ASCA in the energy range 2--10
keV (Della Ceca et al. 1999a find $\Gamma = 1.36 \pm 0.16$ at $S\simeq
6 \times 10^{-14}$ cgs, Ueda et al. 1999 find $\Gamma = 1.6$).  This
may be due to cosmic variance, since we poorly sample the bright end
of the number counts.

For the faint sample we obtained a photon index $\Gamma$ of $1.26 \pm
0.10$ and $N_H = (8.0 \pm 4.0) \times 10^{20}$ cm$^{-2}$ from the fit
using the 0.5--10 keV energy range, and $\Gamma$ of $1.35 \pm 0.20$
with galactic $N_H$ using the 2--10 keV energy range.  The
contribution of the faint sample in the hard band is $ (4.3 \pm
1.0)\times 10^{-12}$ erg cm$^{-2}$ s$^{-1}$ deg$^{-2}$.  In this case
the spectral shape of the faint sample is consistent with the average
spectrum of the XRB, which is around 1.4.  We conclude that the
average spectrum of the detected sources is approaching the average
shape of the hard background (see also Mushotzky et al. 2000).  

The results of the spectral fits are shown in Table 1 and 2.  The new
quantum efficiency file provided with the CXC software was used which
includes the effect of a reduced quantum efficiency at higher energies
caused by the radiation damage.

To test the accuracy of our background subtraction, we perform the
same spectral analysis in two ways.  First, we build a synthetic
spectrum using the software of M. Markevitch (see
http://hea-www.harvard.edu/~maxim/axaf/acisbg/) and we extracted the
background from the same regions used for the extraction of the source
spectra.  In this way we eliminate the uncertainties due to the use of
different extraction regions for the sources and the background.  The
results are the same within a few percent.  Second, we build a
background file summing all the background events found in the annuli
around each source, and scaling the resulting file by the ratio of the
total area af the extraction radius to the total area of the annuli.
In this case too, the spectral fits change by a few percent.

A more detailed view of the spectral properties as a function of the
fluxes comes from the hardness ratio $HR = (H-S)/(H+S)$ where H and S
are the net counts in the hard (2--7 keV) and the soft band (0.5--2
keV), respectively.  The distribution of the hardness ratios as a
function of the count rate in the soft band is shown in Figure 2.
There are 15 sources which are observed only in the hard band plotted
at $HR=1$ (corresponding to 9\% of the total combined sample) and 68
which are observed only in the soft band, and are plotted at $HR=-1$.
When the hardness ratio of the stacked spectrum of the only hard and
only soft sources are computed, we find respectively $HR=0.52\pm 0.07$
and $HR = -0.65\pm 0.05$.  These hardness ratios are plotted as big
asterisks.  It is clear that faint sources are harder than bright
ones, as already shown by the fits of the faint and bright stacked
spectra. 

We note that the stacked spectrum of the 15 sources detected only in
the hard band, is consistent with the typical hardness ratio of the
sources detected in both bands at the count rate of $10^{-4}$ cts/sec
(see asterisks in the upper left of figure 2).  In fact, we detected
soft emission at 3 sigma levels in the stacked spectrum of these 15
sources.  They have been missed in the soft band only because their
emission were just below the detection threshold, while they would
have been detected in a longer exposure.  With this result, we have no
evidence for sources in which the soft emission lower is lower than a
5--10 \% of the energy emitted in the total (0.5--10 keV) band.  This
is expected in synthesis models of the XRB, where a soft component is
considered in obscured AGNs.

\subsection{LogN--LogS and Total Flux from Discrete Sources}

We compute the number counts in the soft and hard bands.  We show in
Figure 3 a comparison of the soft 0.5--2 keV band to the ROSAT results.
We find the Chandra results in excellent agreement with ROSAT in the
region of overlap $S_{min} >10^{-15}$ erg s$^{-1}$ cm$^{-2}$. The
Chandra data extend the results to $2 \times 10^{-16}$ erg s$^{-1}$
cm$^{-2}$.  In Figure 4 we show the LogN--LogS distribution for
sources in the hard band, extending to a flux limit of $2 \times
10^{-15}$ erg s$^{-1}$ cm$^{-2}$, which represents a substantial
improvement over previous missions.  

We performed a maximum likelihood fit with a power law in the range
$2\times 10^{-16}<S<3\times 10^{-14} {\rm cgs}$ for the soft band and
$2\times 10^{-15}<S<5 \times 10^{-14} {\rm cgs}$ in the hard band.
The resulting best fit to the soft LogN--LogS is:
\begin{equation} N(>S) = 370 \Big( {{S}\over
{2\times 10^{-15}}}\Big) ^{-0.85\pm 0.15}\, \, \, {\it sources} deg^{-2} 
\end{equation} 
and to the hard LogN--LogS:
\begin{equation} N(>S) = 1200 \Big( {{S}\over {2\times 10^{-15}}}\Big)
^{-1.0\pm 0.20}\, \,  {\it sources} deg^{-2} 
\end{equation}

The errors on the slope correspond to 2 sigma.  The normalization of
the soft counts is consistent within 1 sigma with the estimate of
Mushotzky et al. (2000).  On the other hand, we find that the previous
Chandra results for the hard counts reported by Mushotzky et
al. (2000), shown as crosses, are higher by 40\% (see \S 5).  From the
maximum likelihood analysis we find that the normalization of the hard
counts is, in fact, more than three sigma lower than the value found
by Mushotzky et al. (2000).  This is partially due to a different
$\Gamma$ used in deriving the converson factors.  If we use an average
$\Gamma = 1.4$, which is appropriate for the faint end of the number
counts, the discrepance in normalization between our hard counts and
that of Mushotzky et al. (2000) is reduced to 3 sigma.  Our hard
counts are consistent with those derived from other deep ACIS-I
pointings, namely the Lynx Field (Stern et al. 2000, in preparation)
and the HDF--N field (Garmire 2000, priv. communication), and in a
completely independent fashion, from the first XMM deep surveys of the
Lockman Hole (Hasinger et al. 2000, A\&A, in press).  In order to
further investigate the difference with Mushotzky et al. results, a
quantitative understanding of the clustering of X--ray sources on
scales $\simeq 5$ arcmin would be needed, together with a knowledge
of the sky coverage of that survey.

The integrated contribution of all sources within the flux range
$10^{-13}$ erg s$^{-1}$ cm$^{-2}$ to $2 \times 10^{-15}$ erg s$^{-1}$
cm$^{-2}$ in the 2-10 keV band is $(1.05 \pm 0.2) \times 10^{-11}$ erg
s$^{-1}$ cm$^{-2}$ deg$^{-2}$.  When this value is added to the
contribution for fluxes $> 10^{-13}$ erg s$^{-1}$ cm$^{-2}$ from ASCA
(Della Ceca et al. 1999b), we have a total contribution of $(1.3 \pm
0.2) \times 10^{-11}$ erg s$^{-1}$ cm$^{-2}$ deg$^{-2}$, which is
close to the value $1.6 \times 10^{-11}$ erg s$^{-1}$ cm$^{-2}$
deg$^{-2}$ from UHURU and HEAO-1 (Marshall et al. 1980).  More recent
values of the 2--10 keV integrated flux from the BeppoSAX and ASCA
surveys (e.g., Vecchi et al. 1999; Ishisaki et al. 1999 and Gendreau
et al. 1995) appear higher by 20--40\%.  The integrated contribution
of all sources in the field plus the bright sources seen by ASCA (data
kindly provided by R. Della Ceca) is shown in Figure 5, together with the
best estimates of the total background.  The inclusion of the ASCA
data at bright fluxes minimizes the effect of cosmic variance.  We
conclude that, given the uncertainty on the value of the total
background, a fraction between $20$\% and $40$\% is still unresolved.

As for the total contribution to the soft X--ray background, we refer
to the 1--2 keV band, following Hasinger et al. (1993, 1998) and
Mushotzky et al. (2000).  We find a contribution of $\simeq 5 \times
10^{-13} $ erg s$^{-1}$ cm$^{-2}$ deg$^{-2}$ from discrete sources for
fluxes lower than $10^{-15}$ erg s$^{-1}$ cm$^{-2}$, corresponding to
$\simeq 11$\% of the unresolved flux ($4.38 \times 10^{-12} $ erg
s$^{-1}$ cm$^{-2}$ deg$^{-2}$).  If this value is summed to the
contribution at higher fluxes (see Hasinger et al. 1998), we end up
with a total contribution of $\simeq 3.5 \times 10^{-12} $ erg
s$^{-1}$ cm$^{-2}$ deg$^{-2}$ for fluxes larger than $2 \times
10^{-16}$ erg s$^{-1}$ cm$^{-2}$, corresponding to 80\% of the
unresolved value.

\subsection{Angular Correlation Function}
  
We have performed an analysis of the clustering of the X--ray sources
in the CDFS using the sources which were identified in both the soft
and hard bands. Since the limited statistics of our sample is expected
to provide a weak clustering signal, we consider the average value of
the two-point angular correlation function, $\bar \omega (\vartheta
)$, defined as the excess of source pairs at separation $\le
\vartheta$ with respect to a random distribution (e.g., Peebles
1980). The random control sample was generated by distributing 50,000
points at random within an area of the sky with the same geometry as
our field and a distribution modulated by the exposure
map. Statistical errors in our value of $\omega(\vartheta)$ are
estimated using the standard bootstrap resampling technique.  The
errors estimated with this technique are those due to statistical
noise and do not include cosmic variance, which can only be assessed
by repeating the analysis over several independent fields.

In Figure 6 we report the results for both the whole source population
and for the subsample of 103 sources with fluxes $> 10^{-15}$ erg
s$^{-1}$ cm$^{-2}$. For reference, we also plot the power--law shape,
$\bar \omega (\vartheta)=(\vartheta_c/\vartheta)^{1-\gamma}$, where
$\gamma=2$ is the slope of the spatial correlation function and
$\vartheta_c=10$ arcsec. Our analysis shows a statistically
significant correlation signal out to scales of about 100 arcsec,
while the noise dominates at larger separations. There is a marginal
indication for the brighter sources to be more clustered at smaller
separations. If confirmed, this would indicate that close pairs tend
to be formed preferentially by bright objects, as is apparent even by
visual inspection of the CDFS.

Vikhlinin \& Forman (1995) analysed the correlation function for the
the angular distribution of sources identified within ROSAT PSPC deep
pointings. Quite remarkably, and despite the different nature of the
sample they used, their correlation function turns out to be
consistent with our Chandra data. Although their much larger
statistics and the wider PSPC field--of--view allowed them to extend
their analysis out to scales $\vartheta \simeq 800$ arcsec, the
exquisite angular resolution achievable with Chandra provides here a
detection of a correlation signal down to scales $\lesssim 10$ arcsec,
about four times smaller than those sampled by Vikhlinin \& Forman
(1995).

We defer to a forthcoming paper a detailed correlation analysis for
the distribution of the Chandra discrete sources and its implication
for models aimed at explaining the nature of X--ray emission from
AGNs.

\section{OPTICAL AND NEAR--IR DATA}

\subsection{Imaging Survey Data}

For identification of sources we have carried out an imaging survey in
3 bands, VRI with the FORS--1 camera at the ANTU telescope (UT--1 of
VLT) (Programme Id 64.O--0621).  We obtained exposures between 4000 to
7000 seconds in 4 adjacent fields of 6.8 x 6.8 arcmin$^2$ covering a
large fraction of the CDFS 16x16 arcmin$^2$ field.  Some shallower
R-band fields of about 1200 seconds each in were obtained with the
FORS imager in 4 additional positions in order to cover the entire
area in the sky swept by different orientations of the Chandra field
of view.  For a few objects we were able to obtain spectroscopic
information using the FORS--1 multislit capabilities in March 2000,
just before the CDFSbecame inaccessible.

Our identification process was complemented with data from the ESO
Imaging Survey (Rengelink et al. 1998) in the J and K bands obtained
with SOFI at the NTT and in the UV bands obtained at the ESO NTT +
SUSI--2. We also used a $30\times 30$ arcmin image in B band obtained
during the commissioning of the WFI camera at the ESO/MPI 2.2 meter
telescope.  The near-IR imaging covers, at present, the central
$9.4\times 9.3$ arcmin of the CDFS.

In the source identification process, we have found a positional
offset of order of 0.3 arcsec and then found the correspondence of
some of the brightest X--ray sources to likely candidates.  We find an
positional offset (-0.2, +1.4) arcsec between the optical and X--ray
data.  Using this correction we find X--ray/optical position
deviations highly concentrated in a radius of rms 0.67 arcsec.  A 2
arcesc correlation radius has been used for further analysis.

\subsection{Optical counterparts of X--ray sources}

Spatial coincidence allows identification of a large fraction of the
sources.  Approximately 10\% of the X--ray sources with more than 10
detected counts have no immediately identifiable optical counterpart
at $m_R < 26$.  This could be due to a flux ratio $S_X/S_{opt}$ larger
than average but still close to the values observed for most of the
sources (see Figure 7).  We find approximately $1/3$ of the sources
are extended, which we associate with galaxies, and $2/3$ of the
sources appear are point like.  If we compare the diagram of X--ray
flux vs R magnitude for ROSAT (Hasinger et al. 1999) and the Chandra
Deep Field South, we find that a large part of the Chandra sources
appear to be consistent with an extension to fainter fluxes of the
ROSAT sources, with an average $S_X/S_{opt} =1$.  However a
significant fraction of the sample ($10\% $) appear to have a very low
$S_X/S_{opt} \leq 1/10$ and is detected only in the soft band at
fluxes $<10^{-15}$ erg s$^{-1}$ cm$^{-2}$.  Most of these objects are
identified as galaxies.  Another subsample of objects that have
$S_X/S_{opt} \geq 10$, appear below fluxes $\simeq 3 \times 10^{-15}$
erg s$^{-1}$ cm$^{-2}$ (soft band).  For these sources (about 20) with
no optical counterpart at $R\simeq 26$, it could be that the
corresponding objects are fainter, thereby increasing the
$S_X/S_{opt}$ to more than $10$, or that they correspond to clusters.

We also plot R-K vs R magnitudes for the Chandra sources (see Figure
8).  The color magnitude plot of the sources in the Chandra Deep Field
South essentially overlaps the same plot with the ROSAT sources and
extends  to fainter magnitudes.  The faintest Chandra sources appear to
have very red colors, R--K $\simeq 5$, with two of them at R--K $\geq
6$, possibly indicating high redshift obscured AGNs.

\subsection{Spectroscopic Information on Selected Sources}

We have obtained spectroscopic data for a dozen optical counterparts
of the X--ray sources using the multislit capability of FORS--1.  The
results are reported in Table 3, where we give the redshift, the
counts in the soft and hard bands, the hardness ratio, the 0.5--2 keV
X--ray flux in erg s$^{-1}$ cm$^{-2}$, the 0.5--2 keV intrinsic
luminosity in units of $10^{43}$ erg s$^{-1}$ cm$^{-2}$, and the
preliminary optical classification based on the width of the emission
lines for Seyferts and QSOs.  The objects described in Table 3 are not
meant to be a well defined sample.  They are just a random sample of
bright CDFS sources for which we obtained the first spectra.  The
optical follow up started October 26 2000.  At present we identify an
elliptical galaxy where we observe an early type continuum with strong
H and K absorption and weak, or no emission lines.

In total we identified 4 QSOs, 5 Seyfert 2 galaxies, 1 elliptical
galaxy and 2 galaxies in an interacting pair.  While we make no claim
for completeness, it is interesting to note that the type of objects
we observe are not extremely different from those we have observed in
the ROSAT surveys.  This can also be shown by plotting the computed
values of $L_X$ against z for the Chandra identifications (Figure 9).
The many objects shown in the diagram are from previous surveys, as
specified by different symbols.  We note that the luminosity of the
objects we have classified as QSOs is of order $1 - 7 \times 10^{43}$
erg s$^{-1}$, while the luminosity of those we classify as Seyfert 2
range between $10^{41}$ and $5 \times 10^{43}$ erg s$^{-1}$ cm$^{-2}$.
The elliptical galaxy has a luminosity of $7 \times 10^{40}$ erg
s$^{-1}$ and the nearby interacting pair galaxies have luminosities of
6 and $9 \times 10^{39}$ erg s$^{-1}$.  None of these objects appear
to exhibit unusual properties.

\section{DISCUSSION}

From the sources detected in the Chandra Deep Field South we resolve
60--80\% of the hard X--ray background.  Their X--ray and optical
properties are consistent with AGNs being the dominant population.

Absorbed AGNs, which have hard X--ray spectra, are missed in shallow
surveys because of absorption dimming. They should be detected
efficiently at low fluxes. The average spectrum of sources detected in
X--ray surveys is expected to become harder with decreasing flux. This
trend, which has already been observed in surveys performed with other
satellites, is shown in Fig. 2.

Although the average spectrum of the sources gets harder with
decreasing flux, only a small fraction ($\sim 9\%$) of the sources are
detected in the hard band and not in the soft band. The hardness ratio
of their stacked spectrum is consistent with the hardness ratios of
the hard sources already detected both in the soft and hard bands,
showing that they may lie just below the soft detection limit.
This might imply that there is not a sizeable population of sources
visible only in the hard band. Indeed, even in highly obscured AGNs,
soft X--ray emission could be produced by partial covering of the
nuclear emission, scattered components, or circumnuclear starbursts
associated with AGN.  Evidence for soft components in obscured AGNs is
commonly observed in good quality X--ray spectra (e.g. Turner et
al. 1997). Furthermore, the analysis of the X--ray color--color plot
for recent ASCA and BeppoSAX surveys (Della Ceca et al. 1999b, Fiore et
al. 2000, Giommi, Perri \& Fiore, 2000) is consistent with this
scenario, with the soft components being 1--10\% of the nuclear
emission.

For about 30\% of the X--ray sources a host galaxy is resolved in the
optical images.  Most of the objects clustered on the lower--left
corner of the soft $S_X$ vs R diagram of Figure 7 are associated with
bright galaxies. These galaxies are overluminous in the R band with
respect to the other X--ray detected sources.  Their X-ray emission
generally comes from the center and is soft.  Many of them are
undetected in the hard band. The upper limits of the hardness ratios
indicate that for the four brightest galaxies the average slope is
$\Gamma \geq 1.5$. For three of these galaxies we have measured
redshifts of z = 0.075, 0.075 and 0.215 with $L_X$ of $9 \times
10^{39}$, $6 \times 10^{39}$ and $7 \times 10^{40}$ erg s$^{-1}$
respectively as given in Table 3.  The remainder of the galaxies are
optically fainter and presumably more distant with luminosity greater
than $10^{41}$ erg s$^{-1}$.  All our soft galaxies fall below the
detection threshold of the deepest ROSAT survey, $10^{-15}$ erg
cm$^{-2}$ s$^{-1}$ (Hasinger et al. 1998), and thus Chandra has opened
up a new parameter space for the study of moderate redshift galaxies
in the X--ray band.

Two sources have been detected in the K but not in the R band (see
Figure 8).  They have R-K $\geq 6$, and can be classified as Extremely
Red Objects (EROs, Elston et al. 1988).  Very red colors would be
produced both by a dust reddened starburst or by an old stellar
population in high redshift galaxies ($z\geq 1$, c.f. Crawford et
al. 2000).  In the first case the X--ray emission could arise from a
surrounding cluster.  Alternatively these objects could be high
redshift obscured AGNs.  The last hypothesis seems more likely for two
objects with a similar color found by ROSAT in the Lockman Hole, based
on their hard X-ray spectrum (Lehmann et al. 2000).  However, the two
EROs in our field are detected only in the soft band, suggesting that
the X--ray emission is likely due to a starburst component or to a
cluster.  Another interesting possibility is that we are observing the
soft scattered component in highly obscured Compton-thick AGNs at high
redshift.

The cumulative number counts in the CDFS are fully consistent with
those of Mushotzky et al. (2000) in the soft band, while in the hard
band we found a surface density lower by $30-40$\% depending on the
average spectral slope $\Gamma$ assumed for the X--ray source
population.  In the soft band the LogN-LogS is within the region
allowed by ROSAT fluctuation analysis (Hasinger et al. 1993), while in
the hard band the BeppoSAX fluctuation analysis (Perri \& Giommi 2000)
does not extend to sufficiently low fluxes to allow a full comparison.
The source counts are in agreement within the errors with the
predictions of AGN synthesis models (Comastri et al. 1995, Gilli et
al. 1999).  Furthermore, the source density observed at low fluxes
seems to favor models where the cosmological evolution of absorbed
AGNs is faster than that of unabsorbed ones  Gilli, Salvati \&
Hasinger (2000).  

\section{CONCLUSIONS}

In the Chandra Deep Field South we have a similarly large survey area
and limiting flux to the Chandra Deep Field North carried out at Penn
State.  Our results are in general agreement with those quoted by this
group who, in their two recent papers (Hornschemeier et al. 2000,
Brandt et al. 2000), concentrated on a limited number ($\sim 10$) of
sources where they had optical identifications.

The limiting flux of the Mushotzky et al. (2000) results is similar
but the area surveyed is about one third of the two Deep Fields.
While we agree on the soft counts, we have a $ \sim$ 40\% lower
normalization of the hard counts, which implies that an important
fraction of the XRB, at least 20\%, has yet to be resolved.

Our data confirm that the XRB is due to the summed contribution of
individual sources.  In the 0.5--2 keV band this result had been
foreshadowed by the Einstein measure of an individual source
contribution of 25\% (Giacconi 1980) and by the ROSAT determination of
a 70\% contribution (Hasinger et al. 1998).  At higher energies (2--10
keV) the contribution to the background from sources was of 25\%
(Cagnoni et al. 1998; Della Ceca et al. 1999a; Ueda et al. 1999;
Giommi, Fiore \& Perri 1998).  Our results bring the individual source
contribution to the XRB in the 2--10 keV range to the level of
60--80\%.  Since there is still a non-negligible fraction of order
20-40\% of the XRB to be resolved, it is important to push the
detection limit at least down to $5\times 10^{-16}$ erg s$^{-1}$
cm$^{-2}$ in the hard band, or even at lower fluxes if the number
counts flatten below $10^{-15}$ erg s$^{-1}$ cm$^{-2}$.

With our increased number of sources we determined that the hardness
ratio increases with fainter limiting flux and the spectrum of the
fainter sources matches the background.  Much of this was expected
from earlier work of Setti and Woltjer (1989), Schmidt and Green
(1986), Madau, Ghisellini \& Fabian (1993), Comastri et al. (1995),
and the ASCA results by Della Ceca et al. (1999a).

In addition, the angular correlation function of the X--ray sources in
our field exhibits significant power on scales of order of the survey
size. The slope and correlation length is consistent with that
observed for galaxies.  This indicates the presence of large--scale
clustering at high redshift, whose spatial extent requires knowing the
redshift distribution of the identified sources, by either
spectroscopic or photometric methods.  Detailed analyses of the
correlation function in Chandra deep pointings will provide important
complementary information on the nature of the discrete X-ray sources
(e.g., Haiman \& Hui 2000, and references therein).

Only 9\% of the sources are detected in the hard band and not in the
soft band.  Most of these sources show some flux in the soft band
below our detection threshold.  As discussed in \S 3.1, the average
hardness ratio of these faint sources is $ HR = 0.52 \pm 0.07 $
consistent with that found for the faint end of the sample.  We
conclude that we find no evidence for sources which are completely
obscured in the X--ray soft band.

Our colour-magnitude and flux--magnitude diagrams follow the trends
well known from the ROSAT data.  A small fraction of the X--ray
sources ($<10\%$) with more than $10$ detected counts are not
immediately identified at R $<$ 26.

A substantial number of nearby galaxies appear in the Chandra survey
at very low $S_x/S_{opt}$ and below the detection threshold of ROSAT.
The X--ray emission from these galaxies is characterized by a soft
spectrum and thus does not contribute significantly to the X--ray
background at high energies. However, their study is of great
intrinsic interest for the understanding of the properties and
evolution of normal and starburst galaxies.

Chandra sensitivity is higher than or equal to that anticipated.
Longer exposures to the confusion limit will still be in a
signal--limited regime.  For a $2 \times 10^6$ s exposure we should
reach $S_{min} \sim 10^{-17}$ erg s$^{-1}$ cm$^{-2}$ in the soft band,
which would allow detection of $L_X = 5 \times 10^{40}$ erg s$^{-1}$
cm$^{-2}$ at $z \sim 0.5$.  This will enable a statistically
significant study of normal and star--forming galaxies to significant
look back times.  Studies of bright QSOs can reach very large
redshifts (5--10) and even observations of low surface brightness
features (e.g., clusters) can be extended to $z \approx 3$.

A deep observation of 385 ksec in the same field is planned during the
guaranteed time of one of us with XMM.  The larger area of XMM will
contribute significantly to the study of the spectra of point sources
and the study of diffuse emission from galaxy clusters.

\acknowledgements

We thank the entire Chandra Team for the high degree of support we
have received in carrying out our observing program.  In particular,
we wish to thank Antonella Fruscione for her constant help in the use
of the CXC software.  We thank also Maxim Markevitch and Pasquale
Mazzotta for discussions.  We thank Roberto Della Ceca for discussions
and for providing the ASCA number counts.  We thank Ingo Lehmann for
his contribution to the spectroscopic redshifts.  G. Hasinger
acknowledges support under DLR grant 50 OR 9908 0.  R. Giacconi and
C. Norman gratefully acknowledge support under NASA grant NAG-8-1527
and NAG-8-1133.

\newpage
 
\begin{table}
\hspace{1.5cm} 
\begin{tabular}{|l|c|c|c|c|c|c|}
\hline
Sample	&  Objects &  Net Counts & $\Gamma$ & $N_H$ & flux (2-10 keV) & $\chi^2/\nu$ \\
 &   &     &  & $10^{19}$ cm$^{-2}$ & cgs deg$^{-2}$ & $ $ \\
\hline
total & 154 & 10100 & $1.61\pm 0.11$& $8$ (galactic) &$
(1.07\pm 0.15) \times 10^{-11}$ & 1.48  \\
bright & 29 & 6050 & $1.70\pm 0.12$ & $8$ (galactic) &$
(6.3\pm 0.9)\times 10^{-12}$& 1.14  \\
faint & 125 & 4050 & $1.35\pm 0.20$ & $8$ (galactic) &$ 
(4.3\pm 1.0)\times 10^{-12}$& 1.21  \\
\hline
\end{tabular}
\caption{Spectral fits on the 2--10 keV energy range.}  
\end{table}

\begin{table}
\hspace{1.5cm} 
\begin{tabular}{|l|c|c|c|c|c|c|}
\hline
Sample	&  Objects &  Net Counts & $\Gamma$ & $N_H$ & flux (0.5-2 keV) & $\chi^2/\nu$ \\
 &   &     &  &  $ 10^{20}$ cm$^{-2}$ & cgs deg$^{-2}$ & $ $ \\
\hline
total & 154 & 10100 & $1.53\pm 0.07$& $6\pm 3$ 
&$ (4.0 \pm 0.3 ) \times 10^{-12}$ & 1.48  \\
bright & 29 & 6050 & $1.71\pm 0.07$ & $7\pm 2$  &$
(2.9 \pm 0.2)\times 10^{-12}$& 0.99  \\
faint & 125 & 4050 & $1.26\pm 0.10$ & $8\pm 4$  &$
(1.0\pm 0.1)\times 10^{-12}$& 1.18  \\
\hline
\end{tabular}
\caption{Spectral fits in the 0.5--10 keV energy range.}  
\end{table}

\begin{table}
\hspace{1.5cm} 
\begin{tabular}{|c|c|c|c|c|c|c|}
\hline

  z & cts & cts & $HR$ &$S_X$ (erg/s/cm$^2$)& $L_X$ ($10^{43}$ erg/s)
   & Opt Id \\ & 0.5-2 keV & 2-7 keV & & 0.5-2 keV & 0.5-2 keV & \\
   \hline

   2.750 &  21.7 &   15.9  &   -0.15 & 6.4 $ 10^{-16}$  &  3.6   &   BALQSO \\
   0.545 & 1162 & 308.5 &   -0.58& 3.3  $  10^{-14} $   &   5.2  &      Sy2 \\
 1.218 & 135.5  & 46.6  & -0.49 & 3.9 $  10^{-15} $     & 3.5    &      QSO \\
0.430 &   44.4 &   0 &  -1.00 & 1.2 $  10^{-15} $   & 0.12   &      Sy2 \\
0.291 & 135.3  & 48.4  & -0.47  &3.9 $  10^{-15} $     & 0.17   &       Sy2 \\
  0.075 &  12.3  & 0  & -1.00  &3.5 $  10^{-16} $  & 0.00088   &Inter. pair \\
 0.075 &   9.1  &  0  & -1.00  &2.3 $  10^{-16} $  & 0.00059   &Inter. pair \\
 0.215 &  9.8  &  0  & -1.00  &3.1 $  10^{-16} $  & 0.0068    &       Ell \\
 0.260 &  11.9  &  0  & -1.00  &3.3 $  10^{-16}$ &  0.011    &        Sy2 \\
1.180 &  55.4  &  21.3  & -0.44  & 1.6 $  10^{-15} $   &   1.3   &      QSO \\
 0.490 &  90.8 &   29.9 & -0.50 & 2.6 $  10^{-15} $   &  0.33   &      Sy2 \\
 1.850 &  114.1 &  34.0  & -0.54 & 3.2 $  10^{-15} $   &   7.5   &      QSO \\
\hline
\end{tabular}
\caption{Sources with optical spectra.}  
\end{table}

\newpage

\begin{figure}
\vspace{6cm} 
{\centerline{\bf Figure 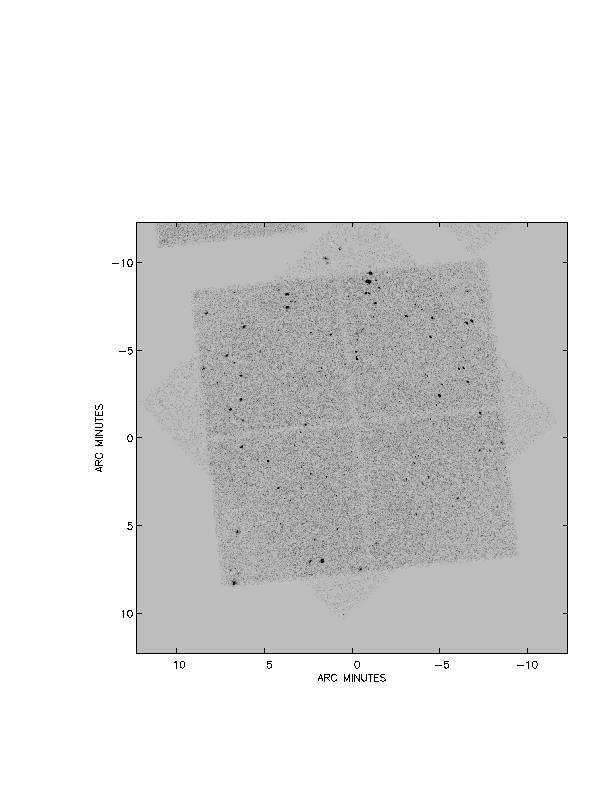}}
\vspace{2cm}
\caption{The Chandra Deep Field in the total band 0.5--10 keV (binned
2x2, pixel=0.984").
\label{fig1}}
\end{figure}

\begin{figure}
\centerline{\psfig{figure=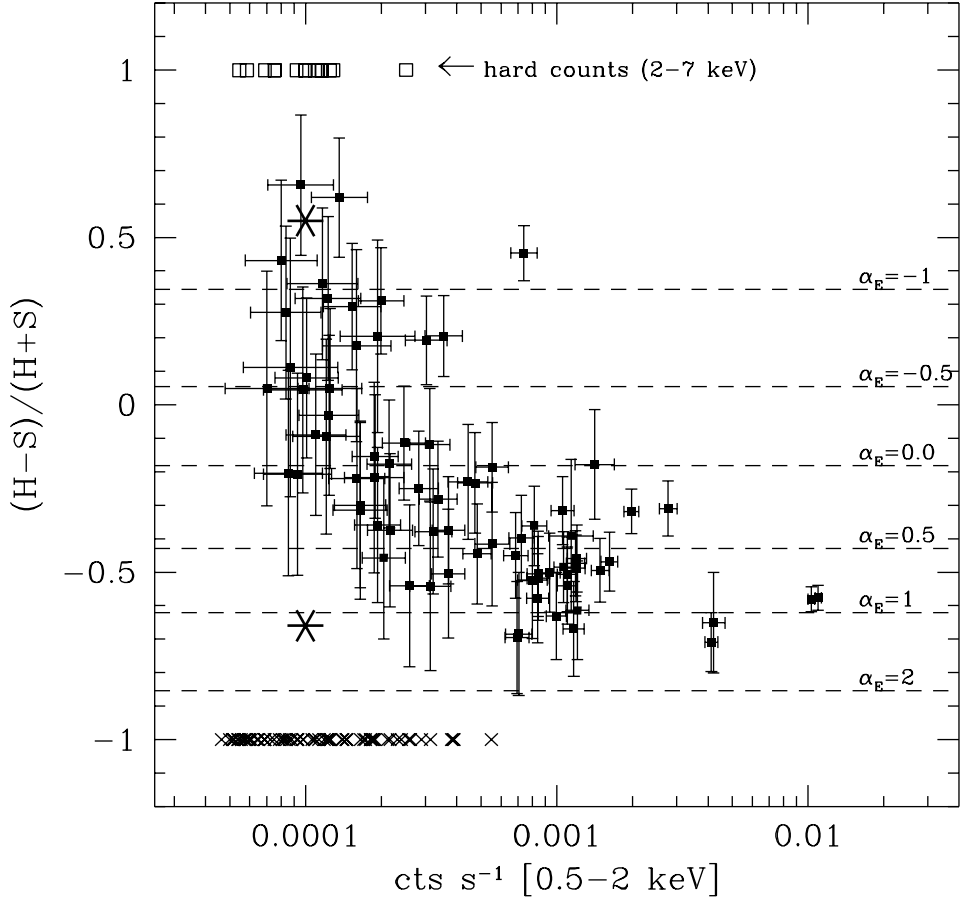,height=6.in}}
\caption{Hardness ratio as a function of the soft count rate for the
sources detected in both bands.  Sources detected only in hard
(HR$=1$) are shown with squares, the ones detected only in the soft
(HR$=-1$) are shown with crosses.  The two asterisks are the hardness
ratio of the stacked spectra of the hard and soft sources.  Dashed
lines are power--law models with different energy index ($\alpha_E$)
computed assuming the galactic value $N_H\simeq 8\times 10^{19}$
cm$^{-2}$\ and convoluted with a mean ACIS-I response matrix at the
aimpoint.
\label{fig2}}
\end{figure}

\begin{figure}
\centerline{\psfig{figure=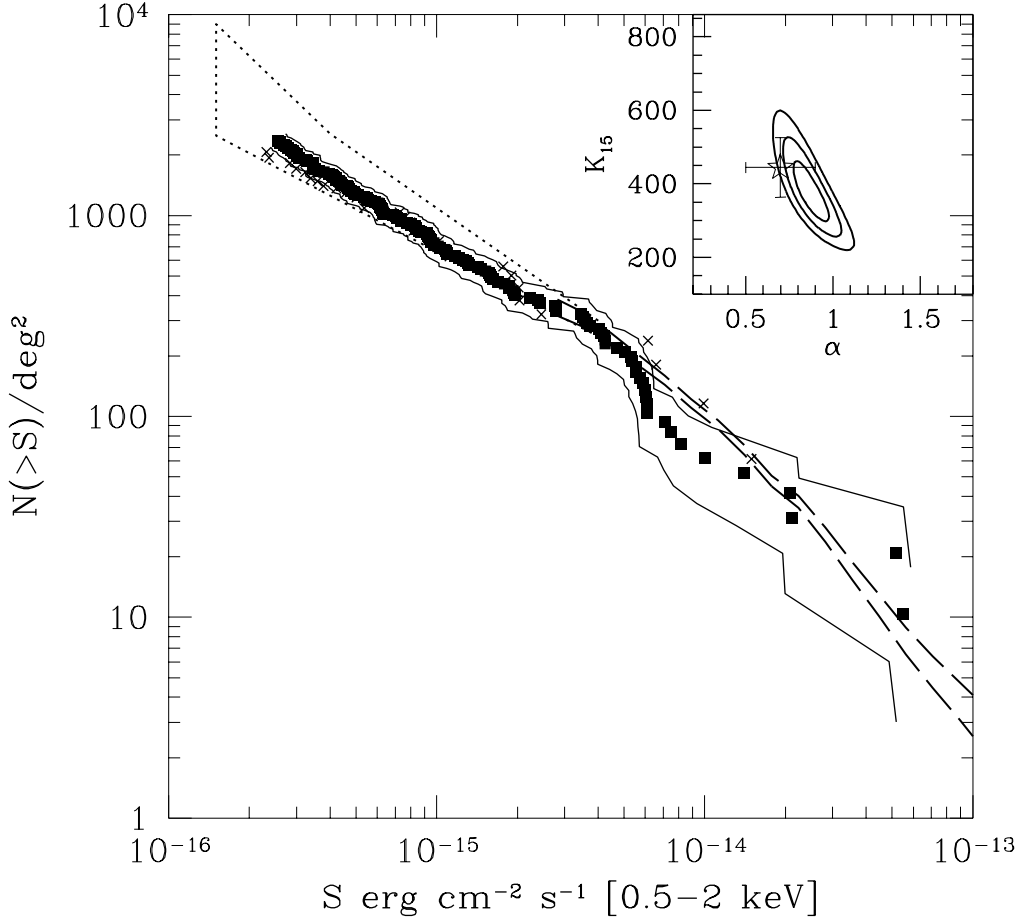,height=6.in}}
\caption{The LogN--LogS in the soft band from the Chandra Deep Field
(filled squares).  The crosses are from Mushotzky et al. (2000).
Dashed lines are the counts from the Lockman Hole (Hasinger et
al. 1998), and the dotted contour is the extrapolation from the
fluctuation analysis in ROSAT data (Hasinger et al. 1993).  The upper
and lower solid lines indicate uncertainties due to the sum of the
Poisson noise (1 $\sigma$) with the uncertainty in the conversion
factor (see text).  The insert shows the maximum likelihood fit to the
parameters in the LogN--LogS fit $ N(>S)=K_{15}(S/ {2 \times
10^{-15}})^{- \alpha}$. The contours correspond to $1 \sigma$, $2
\sigma$ and $3 \sigma$. The star is the fit from Mushotzky et
al. (2000) at $S=2 \times 10^{-15}$ erg s$^{-1}$ cm$^{-2}$; the error
bar is their quoted 68 \% confidence level.  
\label{fig3}}
\end{figure}

\begin{figure}
\centerline{\psfig{figure=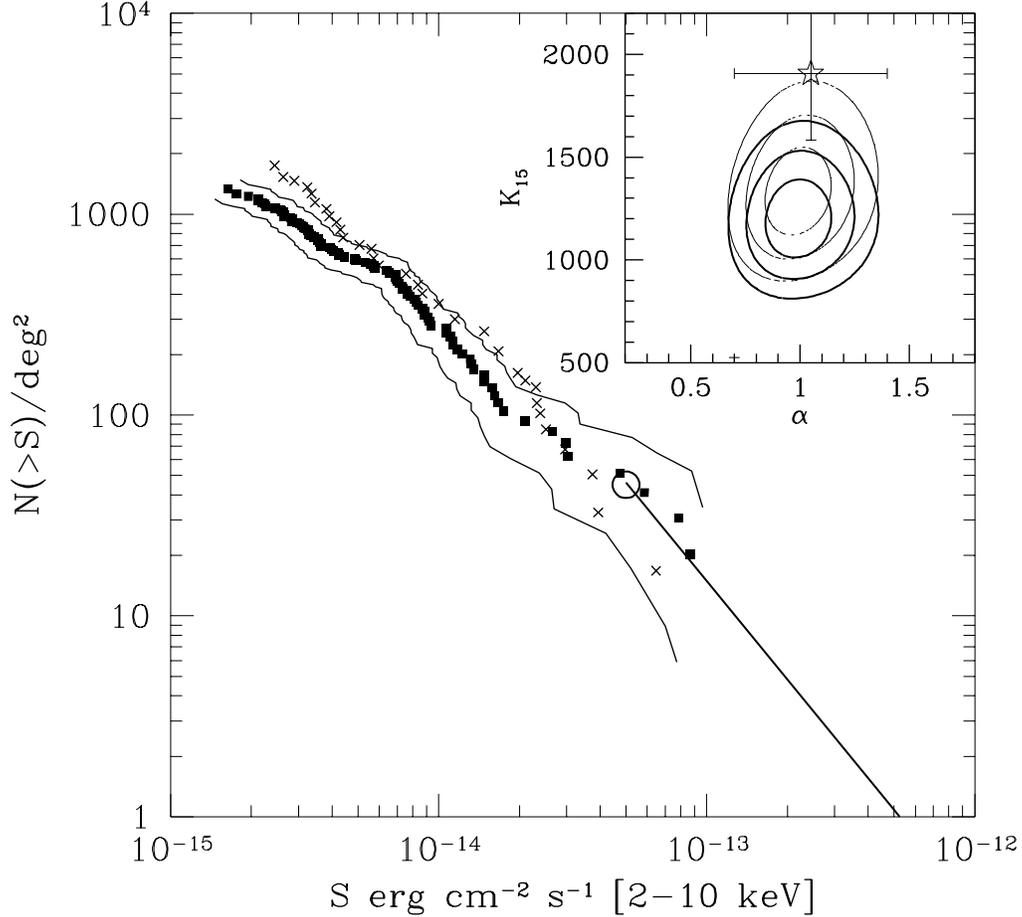,height=6.in}}
\caption{The LogN--LogS in the hard band from the Chandra Deep Field,
symbols as in Figure 3.  The open circle at high fluxes is from ASCA
and Beppo SAX (Giommi et al. 1998, Ueda et al. 1999) and the
continuous line is the fit to the counts from ASCA in the range
$10^{-12}-10^{-13}$ erg cm$^{-2}$ s$^{-1}$ (Della Ceca et al. 1999b).
The upper and lower solid lines indicate uncertainties due to the sum
of the Poisson noise, 1 $\sigma$, with the uncertainty in the
conversion factor (see text).  The insert shows the maximum likelihood
fit to the parameters in the LogN--LogS fit $N(>S)=K_{15}(S/{2 \times
10^{-15}})^{-\alpha}$. The thick contours correspond to $1\sigma$,
$2\sigma$ and $3\sigma$ and use an average photon index of $\Gamma =
1.7$.  The fainter countours use an average photon index of $\Gamma =
1.4$.  The star is the fit from Mushotzky et al. (2000) at $S=2 \times
10^{-15}$ erg s$^{-1}$ cm$^{-2}$; the error bar is their quoted 68 \%
confidence level.
\label{fig4}}
\end{figure}

\begin{figure}
\centerline{\psfig{figure=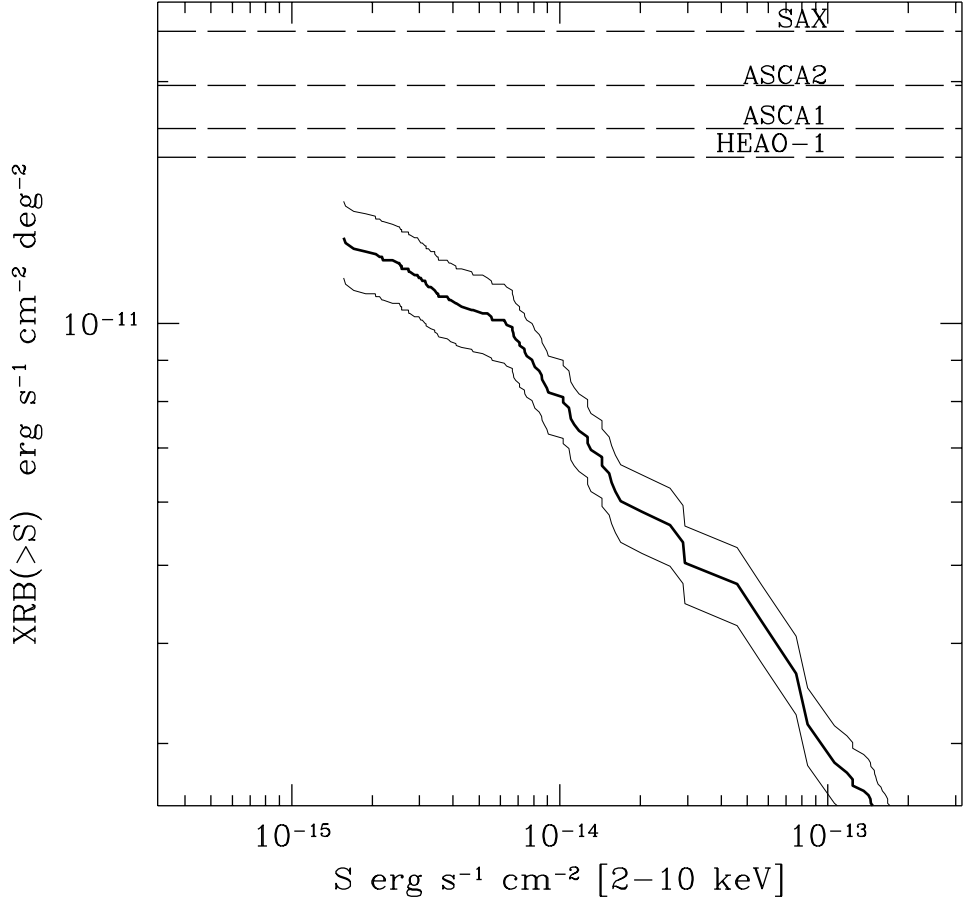,height=6.in}}
\caption{The contribution to the hard X--ray flux density as a
function of the flux of the resolved sources.  The total resolved
contribution is computed from the CDFS sample plus the bright sample
from ASCA at fluxes larger than $\simeq 10^{-13}$ (Della Ceca et
al. 1999b).  The upper dashed lines refer to previous measures of the
hard X--ray background; from bottom to top: Marshall et al. (1980,
HEAO-1), Ueda et al. (1996, ASCA1), Ishisaki (1999, ASCA2), Vecchi et
al. (1999, BeppoSAX). 
\label{fig5}}
\end{figure}

\begin{figure}
\centerline{\psfig{figure=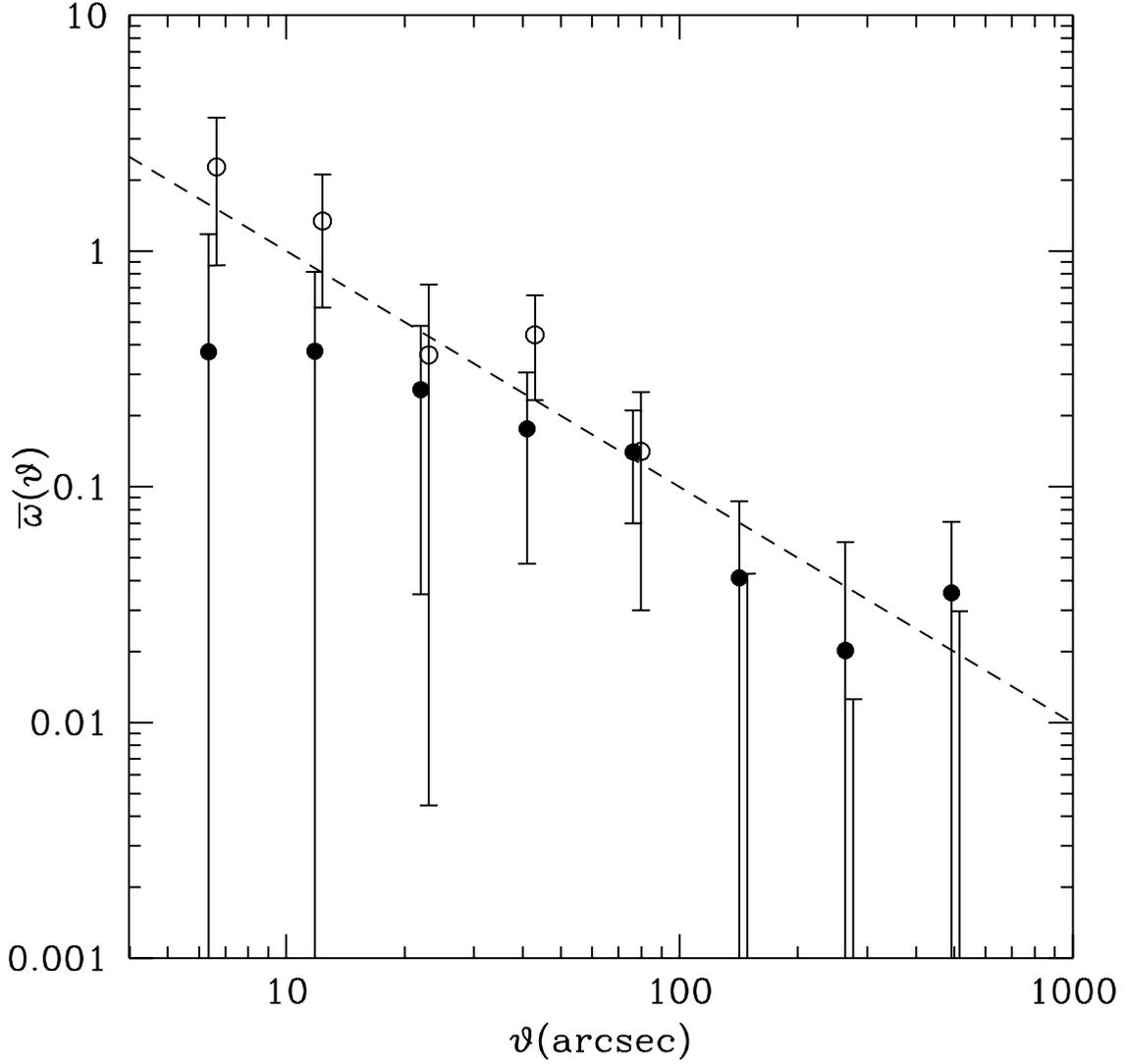,height=6.in}}
\caption{The average correlation function within the angular
separation $\vartheta$ for the projected distribution of discrete
sources identified within the CDFS. Filled circles refer to the whole
sample, while open circles are for the subsample with total flux $S\ge
10^{-15}$ erg s$^{-1}$ cm$^{-2}$. Errorbars are the 1$\sigma$ scatter
estimated over an ensamble of 2000 bootstrap samples (e.g., Barrow et
al. 1984). For reasons of clarity, the open circles have been slightly
shifted along the $x$-axis. The dashed line shows the power law shape,
$\omega(\vartheta)=(\vartheta/\vartheta_c)^{1-\gamma}$, with
$\gamma=2$ and $\vartheta_c=10$ arcsec.
\label{fig6}}
\end{figure}

\begin{figure}
\centerline{\psfig{figure=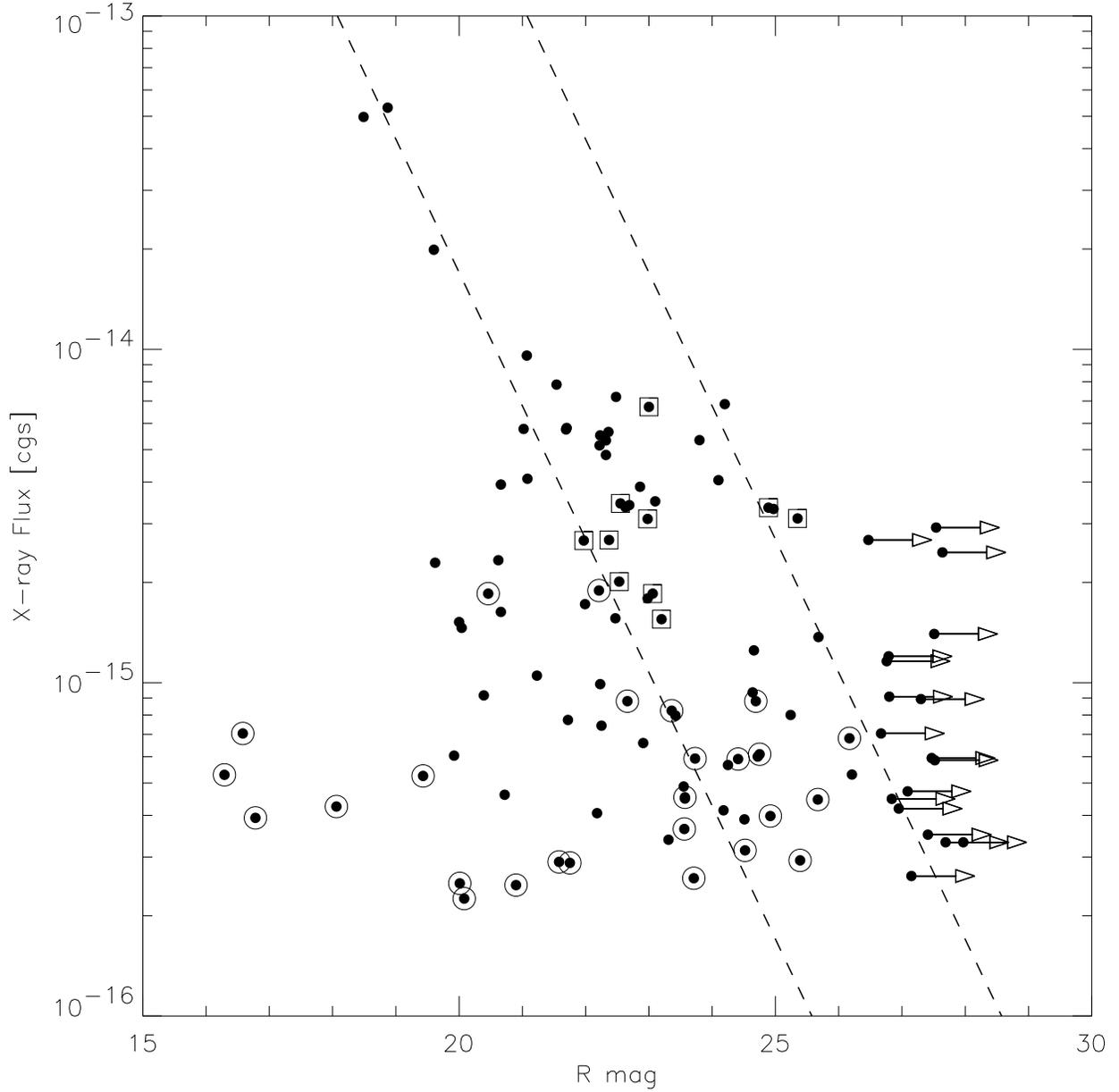}}
\caption{X--ray flux in the soft band versus optical R magnitude for
Chandra sources (dots).  The dotted circles correspond to sources
detected only in the soft band.  The dotted squares correspond to
sources detected only in the hard band.  The left dashed line marks a
constant X--ray to optical flux ratio of 1, which is the best fit
shown by Hasinger et al. (1998) for the ROSAT deep Survey.  The right
dashed line corresponds to an optical flux three magnitudes fainter
($S_X/S_{opt} > 15$).  For the objects without optical counterpart, we
put a $3\sigma$ lower limit to the magnitude (arrows).  Note that
below $F=10^{-15}$ erg s$^{-1}$ cm$^{-2}$, some sources depart more
than three magnitudes from the $S_X/S_{opt}=1$ relation.
\label{fig7}}
\end{figure}

\begin{figure}
\centerline{\psfig{figure=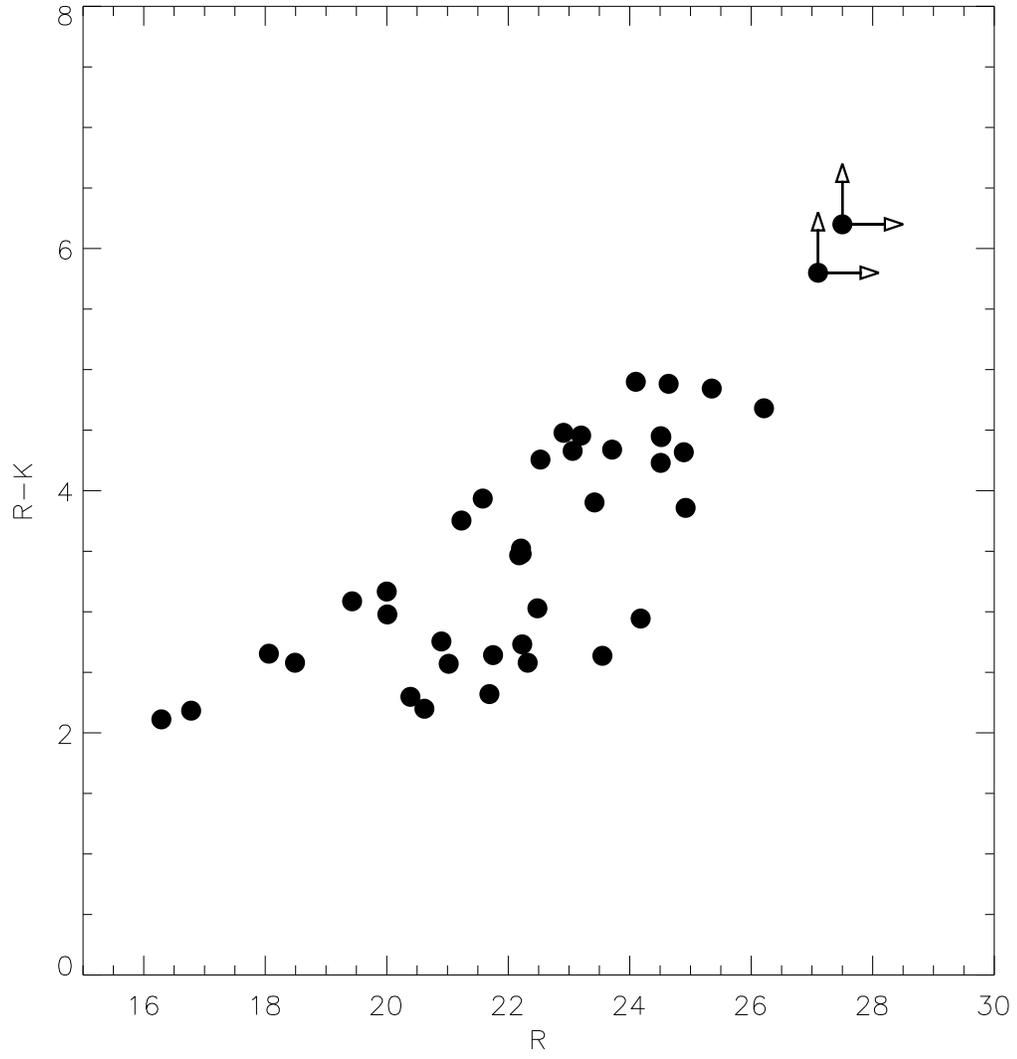}}
\caption{Color-magnitude diagram between R magnitude and R-K color
for Chandra sources.  The two upper right points are X--ray
counterparts detected only in the hard band.  
\label{fig8}}
\end{figure}

\begin{figure}
\centerline{\psfig{figure=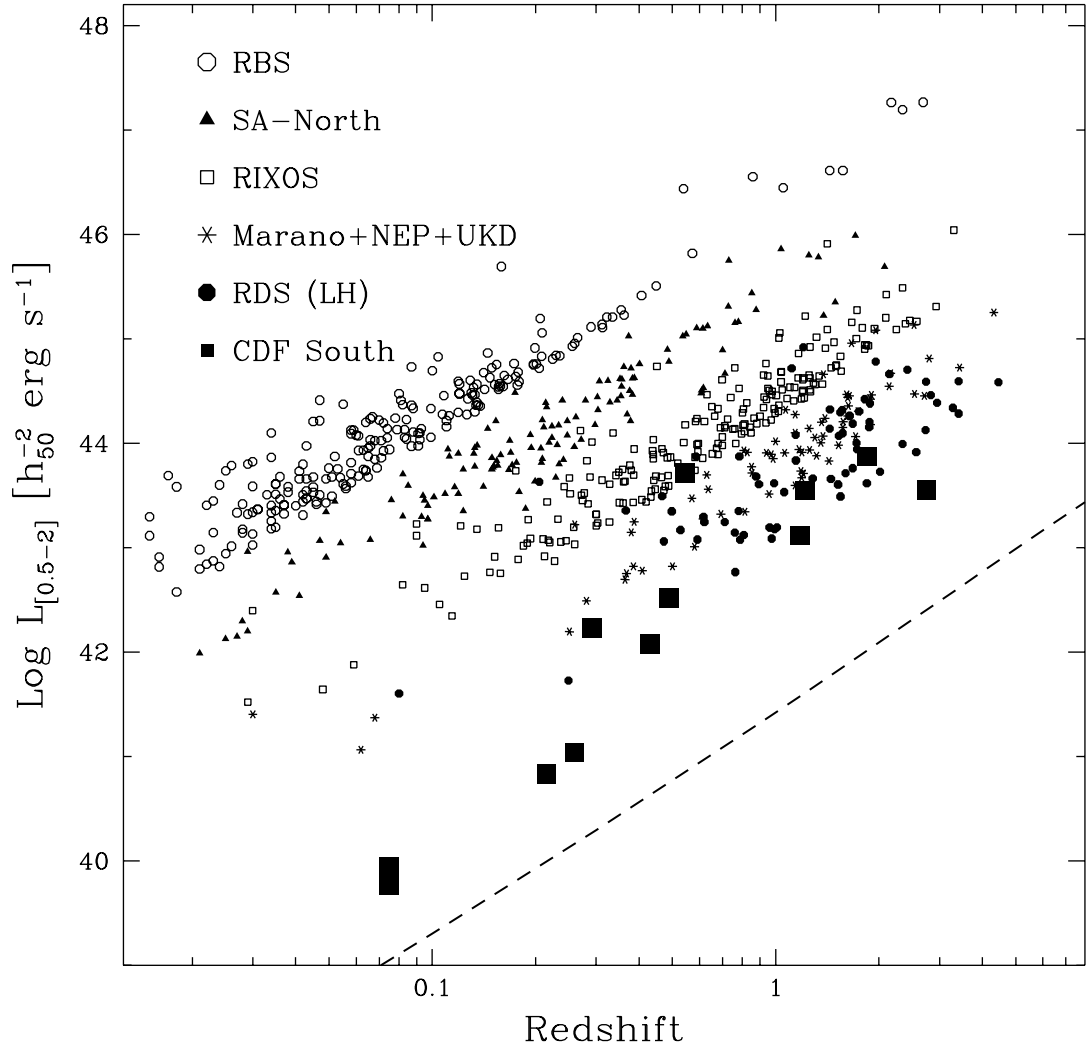,height=6.in}}
\caption{ $L_X$ in the soft band versus redshift for the Chandra
sources (large squares), compared with previous surveys (as shown by
the different symbols).  The dashed line corresponds to a flux limit
of $5 \times 10^{-17}$ erg cm$^{-2}$ s$^{-1}$.
\label{fig9}}
\end{figure}

\end{document}